\documentclass[11pt]{article} 

\usepackage{amsmath,amsfonts,amssymb,amsthm,bbm,stmaryrd}
\usepackage{hyperref}

\newcommand{\be}{\begin{equation}}
\newcommand{\ee}{\end{equation}}
\newcommand{\barray}{\begin{array}}
\newcommand{\earray}{\end{array}}
\newcommand{\bea}{\begin{eqnarray}}
\newcommand{\eea}{\end{eqnarray}}
\newcommand{\bs}{\begin{subequations}}
\newcommand{\es}{\end{subequations}}

\newcommand{\bit}{\begin{itemize}}
\newcommand{\eit}{\end{itemize}}
\newcommand{\bd}{\begin{description}}
\newcommand{\ed}{\end{description}}

\def\nn{\nonumber}

\def\la{\langle}
\def\ra{\rangle}

\newcommand{\p}{\partial}

\newcommand{\R}{\mathbb{R}}

\newcommand{\f}{\frac}

\renewcommand{\d}{\delta}  \newcommand{\eps}{\epsilon} 
 \renewcommand{\th}{\theta}  \newcommand{\vth}{\vartheta} 
    
       \let\r=\rho \let\om=\omega
 \newcommand{\s}{\sigma}      
     
\let\Si=\Sigma \let\Om=\Omega

\def\cF{{\cal F}}

\usepackage{slashed}

\newcommand{\eqons}{\,\hat{=}\,}

\newcommand{\pbi}[1]{\underset{\leftarrow}{#1}}

\newcommand{\sscr}{\scriptscriptstyle\rm}

\usepackage{manfnt}
\reversemarginpar

\newcommand{\Diff}{{\rm{Diff}}}

\DeclareFontFamily{U}{matha}{\hyphenchar\font45}
\DeclareFontShape{U}{matha}{m}{n}{
      <5> <6> <7> <8> <9> <10> gen * matha
      <10.95> matha10 <12> <14.4> <17.28> <20.74> <24.88> matha12
      }{}
\DeclareSymbolFont{matha}{U}{matha}{m}{n}
\DeclareMathSymbol{\oright}       {2}{matha}{"69}

\usepackage{pifont}

\newcommand{\cR}{{\cal R}}
\newcommand{\F}{\cF}

\newcommand{\bac}{\eta}

\newcommand{\jb}{\bar\jmath}
\usepackage{mathrsfs}
\newcommand{\Dd}{{\mathscr{D}}}
\newcommand{\scri}{{\mathscr{I}}}

\begin{document}

\title{\bf General covariance and boundary symmetry algebras}

\author{\Large{Antoine Rignon-Bret and Simone Speziale}
\smallskip \\ 
\small{\it{Aix Marseille Univ., Univ. de Toulon, CNRS, CPT, UMR 7332, 13288 Marseille, France}} }
\date{March 20, 2025}

\maketitle

\begin{abstract}
\noindent 
In general relativity as well as gauge theories, non-trivial symmetries can appear at boundaries. In the presence of radiation some of the symmetries are not Hamiltonian vector fields, hence the definition of charges for the symmetries becomes delicate. It can lead to the problem of field-dependent 
2-cocycles in the charge algebra, as opposed to  the central extensions allowed in standard classical mechanics.  
We show that if the Wald-Zoupas prescription is implemented, its covariance requirement guarantees that
the algebra of Noether currents at the boundary is free of field-dependent 2-cocycles, and its stationarity requirement further removes central extensions.
Therefore the charge algebra admits at most a time-independent field-dependent 2-cocycle, whose existence depends on the boundary conditions. 
We then report on new results for asymptotic symmetries at future null infinity that can be derived with this approach.
\end{abstract}

\section{Introduction}
General relativity shares with gauge theories the property that the only symmetries of the field equations, namely diffeomorphisms, are gauge symmetries: a redundancy in the description and not a physical symmetry relating distinguishable solutions. This is made manifest by Noether's theorem, which proves that all conserved quantities are trivial, and by the symplectic 2-form associated with a slice $\Si$, which is degenerate along every direction corresponding to a diffeomorphism. The situation changes completely in the presence of boundaries. Both Noether charges and the symplectic 2-form are non-vanishing when evaluated on diffeomorphism with non-trivial support at the boundary. The diffeomorphisms, restricted to preserve the boundary conditions that may be present, acquire the interpretation of physical symmetries, whose associated non-trivial charges are surface integrals at the cross-section of $\Si$ with the boundary. Notable examples are the Arnowitt-Deser-Misner (ADM) charges at spatial infinity, which represent different asymptotic inertial observers, and the Bondi-van der Burg-Metzner-Sachs (BMS) charges at (future) null infinity, with an analogue interpretation, and which are particularly relevant for the study of gravitational waves, and play also an important role towards understanding quantum gravity. In all cases the boundary diffeomorphisms form a closed algebra under Lie bracket. One expects that this symmetry algebra be realized by the charges via  the symplectic 2-form,
like in every mechanical system with a phase space description. This is true for the ADM charges, because in this case $\Si$ is a complete Cauchy hypersurface. The BMS case is different because radiation prevents the standard realization of the charge algebra.
Exploring different prescription has led to a literature that abounds which charge algebras afflicted by field-dependent 2-cocycles 
(see e.g. \cite{Barnich:2011mi, Afshar:2015wjm,Barnich:2017ubf, Distler:2018rwu, Barnich:2019vzx,Compere:2020lrt, Geiller:2024amx}),
as opposed to more familiar central extensions. 

We argue that: $(i)$ on-shell field-dependent 2-cocycles are associated with a loss of general covariance, namely a non-trivial dependence on background and non-dynamical structures; hence the corresponding charges may not capture correctly the dynamics of the system, such as misinterpreting mere coordinate effects for `evolution'; $(ii)$ 
if the Wald-Zoupas (WZ) prescription \cite{Wald:1999wa} (or similarly the Hamiltonian construction of \cite{Ashtekar:2024bpi})
can be implemented, then the covariance condition guarantees that the current algebra (namely of the Noether currents on the boundary) is always free of field-dependent 2-cocycles (and furthermore the stationarity condition further prevents central extensions). Hence
the charge algebra (namely of surface charges on a given cross-section $S$ of the boundary) can have at most a time-independent 2-cocycle. 
Removal of the charge cocycle is not guaranteed by the WZ construction, and depends on further considerations such as specific boundary conditions.

We support our proposal discussing various examples relevant to asymptotic symmetries of general relativity at future null infinity,
and reporting on new results whose details appear elsewhere. A WZ potential exists for BMS as well as eBMS \cite{Barnich:2009se,Barnich:2010eb} symmetries, thus the cocycle found in \cite{Barnich:2011mi} can be removed for both current algebras,
but while BMS has a center-less charge algebra, we find for eBMS a residual \emph{time-independent} cocycle.
For the larger groups gBMS \cite{Campiglia:2015yka,Compere:2018ylh} and BMSW \cite{Freidel:2021yqe} finding a WZ potential is harder, however there exists a new symmetry group that incorporates the same $\Diff(S)$ of the previous two groups but with a different action, and for which a WZ potential exists.

\section{Covariance and 2-cocycles}
%
We briefly recall the general formalism and the results of \cite{Iyer:1994ys,Wald:1999wa,Barnich:2001jy,Barnich:2004uw,Barnich:2011mi,Freidel:2021cjp,Odak:2022ndm}. 
The covariant phase space 
is constructed equipping the solution space of a field theory at given boundary conditions with a symplectic 2-form current $\om=\d\th$, where the symplectic potential current $\th$ is read from the on-shell variation of the Lagrangian 4-form, via $\d L\eqons d\th$.
We use the notation of \cite{Harlow:2019yfa,Freidel:2021cjp}
with $\d$ the exterior derivative in field space, $I_V$ the internal product with a vector field $V=\int d^4xV(\phi)\f{\d}{\d\phi}$, and $\d_V = I_V \d+\d I_V$ the field-space Lie derivative. 
Together with their spacetime counterparts $(d,i_v,\pounds_v)$, they define a bi-variational complex with $[d,\d]=0$ (the opposite sign convention is used in \cite{Barnich:2001jy}).
 The field-space vector field corresponding to a diffeomorphism is $V_\xi=\int d^4x \pounds_\xi \phi \f{\d}{\d\phi}$ and we use $\d_{V_\xi}=\d_\xi$ for short.

The Hamiltonian 1-form
of a diffeomorphism for a general covariant theory is a  boundary term on-shell,
\be\label{Ixiom}
-I_\xi\Om_\Si \eqons 
 \d Q_\xi - \cF_\xi.
\ee
Here $\Si$ is a space-like hypersurface that intersects a (time-like or null) boundary of spacetime at the corner $\p\Si$ which we assume compact, and we defined
\be\label{defcf}
\Om_\Si:=\int_\Si\om, \qquad  Q_\xi :=\oint_{\p\Si} q_\xi, \qquad \cF_\xi :=\oint_{\p\Si}i_\xi\th.
\ee
We restrict our discussion to field-independent diffeomorphisms, and refer to \cite{Rignon-Bret:2024gcx} for the general case.\footnote{The symmetry parameters of the BMS, eBMS, gBMS and BMSW groups are all field-independent. It is common to work with symmetry vector fields whose bulk extension is field-dependent through a gauge-fixing condition, e.g. \cite{Barnich:2011mi}. The field dependence of the extension affects the procedure, but only marginally. This detail is discussed in the longer version \cite{Rignon-Bret:2024gcx}.}
If the boundary conditions are strong enough to guarantee that the pull-back at the boundary of $\th$ is field-space exact, namely $\pbi{\th} = \d \ell$, then 
$\pbi\om=0$, $\cF$ is field space exact and all allowed $\xi$'s give rise to Hamiltonian vector fields (HVFs) in the phase space. This is the case for instance for general relativity if the boundary is at spatial infinity, and the resulting generators reproduce the ADM charges \cite{Iyer:1994ys}. The non-trivial situation is when the boundary conditions are `leaky' and allow a flux of radiation. This is the case if $\Si$ is a partial Cauchy slice that extends to future null infinity of an asymptotically flat spacetime, then $\pbi\th=\bar\th + \d\ell$. Now  $\pbi\om\neq0$ and $\cF$ is no longer exact, unless it vanishes if $\xi\in T\p\Si$. 
Hence symmetries that move the corner are not HVFs. This is after all a reasonable situation: symplectic flux means that the system is dissipative, and one cannot expect HVFs to generate motion sensitive to dissipation. 
The problem is that the set of solutions for which $\bar\th=0$ is ambiguous, because one can arbitrarily move terms from $\ell$ to $\bar\th$ and this changes which solutions are singled out. More in general, the field equations only determine the equivalence class $\{\th\sim\bar\th=\th+\d\ell-d\vth\}$ \cite{Jacobson:1993vj,Harlow:2019yfa,Freidel:2020xyx,Odak:2021axr}. The key idea of the Wald-Zoupas paper is to select a split (preferably with $\vth=0$, so that $\om$ is unchanged) so that $\bar\th=0$ for non-radiative solutions. But how to identify gravitational radiation?

At this point, notice that $\bar\th$ is defined only at the boundary. This may look like a minor mathematical point, but it is conceptually profound, because typically it is only thanks to the use of some background structure available at the boundary that one can identify gravitational radiation. For instance, the Bondi news  at future null infinity  relies on a choice of conformal compactification \cite{Newman:1968uj,Geroch:1977jn}. One can then require that $\bar\th$ vanishes on spacetimes with vanishing news, which we will refer to as `stationarity condition'.
Another background structure that is often used is a foliation of $\scri$, that allows one to talk about the asymptotic shear of gravitational waves.
One has to make sure that the use of any background field does not contaminate the physics, whence the importance of requiring covariance.
Denoting generically $\eta$ any background field, the precise mathematical property we require is
$\varphi^*\bar\th [\phi, \d \phi; \bac] = \bar\th [\varphi^* \phi, \d( \varphi^* \phi); \bac]$ for a diffeomorphism $\varphi$ that corresponds to an asymptotic symmetry.\footnote{In the BMS case this requirement may be supplemented requiring independence under arbitrary conformal transformation, but this can also be studied via covariance, just using the bigger BMSW group as an auxiliary tool \cite{Odak:2022ndm,Rignon-Bret:2024gcx}.}
At the linearized level, this means that the Lie derivative in field space and spacetime coincide, namely
\be\label{covth}
\d_\xi \bar\th = \pounds_\xi \bar\th.
\ee
This equation has the important consequence that
\be\label{Ixiom2}
-I_\xi\pbi{\om} = -I_\xi{\bar\th} = -\d_\xi {\bar\th} + \d I_\xi{\bar\th} = \d I_\xi{\bar\th} -d i_\xi{\bar\th}.
\ee
If the symmetry vector fields are tangent to the boundary, Noether's theorem applied to a covariant Lagrangian implies that $I_\xi\bar\th$ is on-shell exact \cite{Iyer:1994ys}. We can thus define a `bi-exact' quantity
\be\label{qWZdef}
\d {d \bar q}_\xi :\!\!\!\eqons -I_\xi\pbi{\om}+d i_\xi{\bar\th} =  \d I_\xi{\bar\th}.
\ee
This is the Wald-Zoupas prescription for the charges \cite{Wald:1999wa} (see also \cite{Grant:2021sxk,Odak:2022ndm,Ashtekar:2024bpi} for more recent reviews and further discussions).
It satisfies the flux-balance law
\be\label{WZflux}
{d \bar q}_\xi \eqons \jb_\xi:= I_\xi {\bar\th}.
\ee
The ambiguity of adding a field-space constant in going from \eqref{qWZdef} to \eqref{WZflux} can be removed requiring the stationarity condition for $\jb_\xi$ as well. We denote $\jb_\xi$ the Noether current, and $q_\xi$ the charge aspect.\footnote{Some literature e.g. \cite{Barnich:2019vzx} uses the term charge current to refer to what we call the aspects, namely the 2-forms. In this paper we will always use current to refer to the 3-form Noether currents.}

Equation \eqref{WZflux} identifies $q_\xi$ only up to a closed 2-form, hence $Q_\xi$ only up to time-independent terms. 
On the other hand, $d\om\eqons 0$ guarantees that 
\be\label{qWZSi}
\d \bar Q_\xi \eqons -I_\xi\Om_\Si +\oint_S i_\xi {\bar\th},
\ee
providing the relation between the Wald-Zoupas charges and Hamiltonian generators on $\Si$. In particular, the WZ charges provide canonical generators of symmetry vector fields tangent to the corner, whereas for the remaining vector fields tangent to the boundary but not the corner they provide canonical generator only for arbitrary perturbations around the stationary solutions. If $\Om_\Si$ is fixed, as in the original WZ prescription, the relation \eqref{qWZSi} removes the ambiguity of adding time-independent terms, up to a new ambiguity of field-space constants. This final ambiguity can be removed requiring that the charges vanish on a reference solution, e.g. Minkowski. But if one allows non-trivial corner terms $\vth$, as required for instance by enlargements of the BMS group, it is possible to use them to modify $\Om_\Si$ while keeping the same $\bar\th$. In this case, \eqref{qWZSi} is not enough to remove the ambiguity of adding time-independent constants to the charges, and a different prescription is needed, as we explain below.

All we discussed so far is well-known. The new result we would like to point out is that there is another, very important property that covariance guarantees: the correct realization of the symmetry algebra in the phase space.
Previous results have in fact shown that for a general split \cite{Barnich:2011mi,Barnich:2019vzx,Chandrasekaran:2020wwn,Freidel:2021cjp,Wieland:2021eth}
\begin{align}\label{chargealgebra}
-I_\chi I_\xi\Om_\Si &\eqons \d_\chi Q_\xi -I_\chi\cF_\xi = Q_{[\xi,\chi]} + K_{(\xi,\chi)} +I_\xi\cF_\chi-I_\chi\cF_\xi,
\end{align}
where $K_{(\chi,\xi)}$ is a 2-cocycle, namely it satisfies the Jacobi identity, and can be \emph{field-dependent}, namely $\d K\neq0$. This should be contrasted with the situation familiar from non-dissipative systems in classical mechanics, where all symmetries correspond to Hamiltonian vector fields, and $\cF$ vanishes. There one can prove that $K$ is at most  a central extension, namely $\d K=0$. 
This turns out not to be true for a general covariant theory with background structures at the boundaries, because $K$ can be field-dependent even at points in the phase space where $\cF$ vanishes and the symmetries are locally HVFs \cite{Barnich:2011mi,Barnich:2019vzx,Chandrasekaran:2020wwn,Freidel:2021cjp}. An explicit example comes from the calculations done in \cite{Barnich:2011mi,Barnich:2019vzx} concerning  gravitational radiation at future null infinity. The split there chosen vanishes whenever the time derivative of the shear vanishes, hence the charges associated to that split are canonical generators around those solutions. However, the charge algebra has a field-dependent 2-cocycle if one uses a conformal factor that is not a round sphere!
We now show that this unusual situation is the consequence of having lost covariance.

To understand the effects of covariance, observe that
$\d_\xi-\pounds_\xi$ measures the contribution to the transformation laws that come from non-dynamical fields. If the charges are covariant, then the only source of this operator should be the symmetry representatives themselves, namely 
\be\label{covcurrent}
(\d_\chi-\pounds_\chi)\bar\jmath_\xi
= \bar\jmath_{[\xi,\chi]} . 
\ee
Thanks to the commutator $[\d_\chi -\pounds_\chi, I_\xi ] = I_{[\xi, \chi]}$, it is immediate to see that this property holds iff $\bar\th$ satisfies \eqref{covth}, 
namely it is covariant at the linearized level.
From this property we immediately find that
$
-I_\chi I_\xi \pbi\om  \eqons \jb_{[\xi, \chi]} + d(i_\chi I_\xi{\bar\th} - i_\xi I_\chi{\bar\th} ).
$
Following Barnich and Troessaert we can then define a \emph{current bracket} subtracting the flux, namely
\begin{align}\label{currentalgebra}
\{\jb_\xi,\jb_\chi\}_*&:=-I_\chi I_\xi \pbi\om +d(i_\xi I_\chi{\bar\th} - i_\chi I_\xi{\bar\th}) = (\d_\chi -\pounds_\chi) \jb_\xi = \jb_{[ \xi,\chi]}.
\end{align}
This is our first result: \emph{the current algebra associated with a covariant split is free of 2-cocycles.}
In fact, it is also free of central extensions. This may look surprising at first, but a moment of reflection shows that it
is a consequence of the stationarity requirement. Without it, we could have added a field-space constant $-a_\xi$ to the RHS of \eqref{WZflux}, which in turn would have added a central extension $-a_{[\chi,\xi]}$. 
 
Coming to the surface charges, \eqref{currentalgebra} implies
\be
(\d_\chi-\pounds_\chi) \bar q_\xi = \bar q_{[\xi,\chi]} + k_{(\xi,\chi)}, \qquad d k_{(\xi,\chi)}=0.
\ee
Integrating over a cross-section and using \eqref{WZflux}, we find the Barnich-Troessaert bracket
\begin{align}\label{BTbracket}
\{ Q_{\xi}, Q_\chi \}_* &:= \d_\chi Q_\xi - I_\xi {\F}_\chi = -I_\chi I_\xi \Om_\Si +I_\chi \F_\xi - I_\xi \F_\chi \eqons Q_{[\xi,\chi]} + K_{(\xi,\chi)},
\end{align}
where $K_{(\xi,\chi)}=\oint k_{(\xi,\chi)}$ and
\be
\pounds_n K_{(\xi,\chi)}=0
\ee
since $k_{(\xi,\chi)}$ is closed.  
This is our second result: \emph{
the only possible cocycle in the charge algebra with a covariant symplectic potential is a closed 2-form}.
Equivalently, a time-independent 2-form with respect to the flow of the cross-sections.
We conclude that a covariant symplectic potential is compatible with a time-independent 2-cocycle in the charge algebra. 
Knowing if it is present or not, and whether it is a central extension, requires additional information.

Our analysis shows the importance of covariance of the symplectic potential as a 3-form. Covariance of the 3-form up to total divergences is not enough, one reason being that typically the anomaly operator and covariant derivative do not commute.

\section{Case studies}
%
Simple examples of boundaries for which 
the WZ conditions are satisfied are spatial infinity \cite{Iyer:1994ys} and null hypersurfaces in physical spacetimes \cite{Chandrasekaran:2018aop,Shi:2020csw,Ashtekar:2021kqj,Odak:2023pga,Chandrasekaran:2023vzb}.
In all cases one can obtain charges that are manifestly covariant in the WZ sense, hence both the current and the charge algebras are center-less.
The non-trivial cases were our approach brings new result is future null infinity with various boundary conditions.

\medskip

\emph{BMS symmetries}. This was the main goal of the WZ paper, and it was found that there is a unique symplectic potential (with $\vth=0$)
that satisfies the covariance and stationarity requirements. It can be written as
\be\label{thBMS}
\th^{\sscr BMS} = -\f1{16\pi G} N_{ab}\d \s^{ab}\eps_\scri,
\ee
where $\s_{ab}$ is the shear of a foliation Lie-dragged by the normal vector $n$ and  in this case  the news tensor is
$N_{ab}= 2\pounds_n \s_{ab} -\r_{\la ab\ra}$ \cite{Dray:1984rfa}, where $\r_{ab}$ is Geroch's tensor and $\la,\ra$ stands for trace-free.
The corresponding charges coincide respectively with Geroch's super-momentum \cite{Geroch:1977jn} and with Dray-Streubel's Lorentz charges (boosts and angular momentum) \cite{Dray:1984rfa}. 
Their flux was proven in \cite{Dray:1984gz} to match the Ashtekar-Streubel flux \cite{Ashtekar:1981bq}. 

From our general argument it follows that the algebra of BMS fluxes is center-less between any two cross-sections of $\scri$. In this case also the charge algebra is center-less because there are no covariant and conformally invariant corner terms that are also time independent. The explicit calculations reported in \cite{Rignon-Bret:2024gcx} are instructive, and highlight that the key property to have covariance in any conformal frame is \cite{Geroch:1977jn} 
\be\label{lierho}
\pounds_\xi \r_{ab}=-2D_aD_b\dot f,
\ee
where $\dot f=\pounds_nf$ ($n$ is the normal to $\scri$) and $f$ is the vertical part of the symmetry vector field $\xi$.

The 2-cocycle found in \cite{Barnich:2011mi} for conformal frames that are not round spheres is thus entirely due to a `bad split', in which the contribution from Geroch's tensor was dropped in the non-integrable term, and accordingly also in the charges. It is a good split if the frames are restricted to be round spheres, and indeed the 2-cocycle vanishes in these frames. The reason for this difference is that $\pounds_n \s=0$ is a covariant statement only if the conformal compactification uses a round sphere, and not otherwise.

\medskip

\emph{eBMS symmetries}. In the eBMS case the symmetry vector fields are allowed to be singular, namely to be non-globally defined conformal Killing vectors of the cross-sections. As a consequence, the topology arguments used by Geroch to prove unicity and universality of his tensor are lost. Hence also \eqref{lierho} is lost, and replaced by 
\be\label{drho}
\pounds_\xi \r_{ab}= -2D_aD_b\dot f +\d_\xi\r_{ab}
\ee
with a non-zero $\d_\xi\r_{ab}$ \cite{Compere:2016jwb}.
This fact has a dramatic effect for the flux of the BMS charges, which is \emph{no longer zero even if the news vanishes}. For the same reason, the symplectic potential \eqref{thBMS} associated with this split picks up an extra term, given by $-\f1{16\pi}\s^{ab}\d\r_{ab}$. It is invariant under linearized conformal transformations, hence satisfies linearized covariance, and there is no cocycle in the algebra of Noether currents. The contribution from $\r_{ab}$ was not included in \cite{Barnich:2011mi},
and this explains the cocycle there found. The crucial observation is that $\d_\xi\r_{ab}\neq 0$  is \emph{inhomogeneous}, hence the cocycle appears also in a frame with $\r_{ab}=0$. Geroch's tensor thus removes the cocycle \cite{Barnich:2011mi} in both BMS and eBMS.

This said, the extra term is \emph{not} invariant under finite conformal transformations, nor stationary. 
Hence even though including it removes the cocycle, the symplectic potential is still not WZ. Remarkably, it is possible to solve this problem, but it requires a modifucation of the symplectic 2-form, with a corner term improvement. 
The Ashtekar-Streubel radiative phase space at $\scri$ contains also non-radiative configurations (`vacua') \cite{Ashtekar:1981bq,Ashtekar:1981hw,Ashtekar:2024bpi}. Picking a foliation $u$ of $\scri$ given by say an affine parameter, we can parametrize the non-radiative configurations in terms of a `bad cut' $u_0(x^A)$, a quantity that can be recognized as the `super-translation field' of \cite{Compere:2018ylh}.
We then define \cite{Rignon-Bret:2024gcx}
\begin{align}\label{theBMS}
\th^{\sscr eBMS} &:= \th^{\sscr BMS}-d\vth = - \frac{1}{16 \pi} N_{ab} \left(\d \s^{ab} +\f12 (u - u_0) \d \rho^{ab}\right) \eps_\scri,
\end{align}
where
\be
\vth = \f1{16\pi} (u-u_0)\left( \s^{ab} -\f14(u-u_0)\r^{ab} \right)\d\r_{ab} \eps_S.
\ee
The new symplectic potential is  WZ for eBMS, hence the current algebra is center-less.
$\th^{\sscr eBMS}$ provides a local version (in the sense of being a 3-form) of the one proposed in \cite{Campiglia:2020qvc,Donnay:2022hkf}. 
Our results thus generalize these references providing Noether currents whose algebra is center-less
when integrated between arbitrary cuts of $\scri$, and not only over all of $\scri$,
 hence independently of fall-off conditions at $u\to\pm\infty$.

The charge algebra on the other hand has a time-independent cocycle, such that super-rotations charges act covariantly only on global translations. We were not able to identify a charge prescritpion that is fully covariant. The time-independent terms that are not universal and enter the cocycle are $\r_{ab}$ and $u_0$ \cite{Rignon-Bret:2024gcx}. 

\medskip

\emph{gBMS, BMSW and RBS symmetries}. 
For the for weaker fall off conditions leading to gBMS \cite{Campiglia:2015yka,Compere:2018ylh} and BMSW \cite{Freidel:2021yqe} it is harder to satisfy the covariance condition \cite{AS2}. The difficulty comes from a non-trivial spatial dependence of $\dot f$
that prevents commutation between Lie derivatives and pull-back on $\scri$. While we are still working on this problem, one possibility to avoid it is to restrict attention to symmetries preserving the normal without conformal rescaling. This can be understood as a change of universal structure to preserving a late-time preferred frame, which we refer to as `rest-frame Bondi spheres' (RBS). Its symmetry group consists of arbitrary diffeomorphisms of the spheres, super-translations, and a global dilation \cite{AS2}:
\be
	{\rm G}^{\sscr RBS}:=	(\Diff(S) \ltimes \mathbb{R}) \ltimes \mathbb{R}^{S}.
	\label{GRBS}
\ee
When the Bondi condition holds, it is a subgroup of BMSW. In spite of having the same symmetries of the gBMS group (plus the global dilation), it does not contain gBMS as a group, because the structure constants are different. It only shares with gBMS the subgroup SDiff$(S) \ltimes \mathbb{R}^{S}$ of divergence-free diffeomorphisms on the sphere and super-translations. For the RBS group we found a one-parameter family of WZ symplectic potentials, given by \cite{AS2}
\begin{align}\label{RBSWZth}
& {\theta}^{\sscr RBS} = -\frac{1}{16 \pi G} \bigg[\bigg(N_{ab} \d \mathcal{S}^{ab} \eps_I  + \f14 (u - u_0) (\cR N^{ab} - 2\mathcal{D}^a \mathcal{D}^c N^{bc}) \d \hat{g}_{ab} 
\bigg)\eps_I \\ \nn
	& \hspace{1.5cm} +\bigg( \f14(u - u_0)( N_{ab}N^{ab}+ 6 \mathcal{D}_a \mathcal{D}^bN^{ab} )+ \beta \,  \mathcal{S}_{ab} N^{ab}\bigg)\d \eps_I\bigg].
\end{align}
Here 
$
	\mathcal{D}_a := \Dd_a + \p_a u_0 \pounds_n
$
is a `Fourier' covariant derivative (since the charge it sees is the frequency of the field), and $ \mathcal{S}_{ab}= \sigma_{ab} + D_{<a} D_{b>} u_0$ is the covariant or relative shear \cite{Compere:2018ylh,Ashtekar:2024bpi,Ashtekar:1981hw}. The free parameter $\beta\in\R$ corresponds to one of the two free parameters considered in \cite{Compere:2019gft}.\footnote{The other one is forbidden by covariance, a result that resonates with the lack of covariance found in \cite{Chen:2022fbu}.}
The Noether current algebra obtained from \eqref{RBSWZth} provides a center-less representation of the RBS symmetry algebra.

\medskip

 \emph{Corner charges defined as currents}. We have seen that the WZ prescription only guarantees that the Noether currents provide a center-less representation of the algebra, whereas for the corner charges, a time-independent but field-dependent cocycle can still be present. If this is the case, the background-independence hence physical meaning of the corner charges is unclear. 
A possible (trivial) way out would be to define the charges as differences with respect to a subtraction point, say in the distant future where spacetime is assumed to settle down to a special solution. For instance if we assume that spacetime settles down to Minkowski spacetime at $u \rightarrow + \infty$, then we can define the corner Noether charge any cross section $S$ of $\scri^+$ as the integral of the current between $\scri^+_+$ and $S$. By doing this, the charge vanishes automatically at any cross section $S$ in Minkowski spacetime, since the current vanishes in this case, and the charge algebra is free of cocycle. 

\section{Conclusions}
Noether's theorem provides a beautiful approach to the problem of observables in general relativity.
However, the identification of the charges and their currents has to be done with care. Choices have to be made, and it was the key message of \cite{Wald:1999wa} (and previously of \cite{Geroch:1977jn,Ashtekar:1981bq}) that these should be guided by the physical considerations of stationarity and covariance. We have found a new implication of these guiding principles: they guarantee that the symmetry algebra is also realized correctly in the phase space, 
at least at the level of the currents. `Integrating' the currents to get the charges is a non-trivial step, and may introduce again field-dependent cocycles, although limited to be time-independent. As applications of our result, we have reported that identifying a WZ symplectic potential to prescribe charges removes the 2-cocycle found in \cite{Barnich:2011mi,Barnich:2019vzx}, provides a local  extension of the symplectic potential for eBMS found in \cite{Campiglia:2020qvc} and \cite{Donnay:2022hkf},
and suggests a new avenue to include general $\Diff(S)$ transformations. 
Details and the explicit expression for the charges can be found in \cite{Rignon-Bret:2024gcx,AS2}. 

The key message that a 2-cocycle in the charge algebra stems from a loss of covariance at some level was already explicit in the formulas of \cite{Chandrasekaran:2020wwn,Freidel:2021yqe,Freidel:2021cjp}. We have sharpened it identifying the precise physical origin of the loss of covariance from the conformal and foliation dependence, and more importantly showing the implications of the Wald-Zoupas requirements. We have explained the role of the stationarity condition in removing central extensions in the current algebra, pointed out the difference between current and charge algebras, and the role of boundary conditions in the charge cocycle. Absorbing the charge cocycle in a redefinition of the bracket, as considered in \cite{Freidel:2021cjp}, is not viable because it hides the fact that the charges don't generate the symmetries for the stationary configurations.

Another conclusion that can be drawn from this analysis is that the current algebra is on more solid grounds than the charge algebra. This resonates with the proposal of \cite{Ashtekar:2024bpi} (see also \cite{Ashtekar:1981bq,Ashtekar:1981hw}) to identify the Hamiltonian generators of the asymptotic symmetries as fluxes on the local radiative phase space on regions of $\scri$, rather than as surface charges on $\Si$.

Finally, while we focused our applications to asymptotic symmetries at $\scri$, we hope that the general result on covariance and symmetry algebras will be relevant for other gravitational systems.

\providecommand{\href}[2]{#2}\begingroup\raggedright\endgroup

\end{document}